\documentclass[aps,onecolumn,nofootinbib,notitlepage,superscriptaddress]{revtex4}
\usepackage{graphicx}
\usepackage{dcolumn}
\usepackage{bm}
\usepackage{amsmath,amssymb}
\usepackage{float}
\usepackage{multirow}
\usepackage{slashed}
\usepackage{xcolor}
\usepackage{physics}
\usepackage{multirow}
\usepackage[colorlinks=true, pdfstartview=FitV, bookmarks=true, bookmarksnumbered=true, breaklinks]{hyperref}
\usepackage{mathtools,braket}
\usepackage{soul}
\usepackage{lipsum}  
\usepackage{color}
\definecolor{blue}{rgb}{0.0, 0.0, 1.0}
\definecolor{red}{rgb}{1.0, 0.0, 0.0}
\definecolor{royalblue}{rgb}{0.0, 0.14, 0.4}
\hypersetup{linkcolor=royalblue, citecolor=blue, urlcolor=royalblue}

\usepackage{hyperref}
\hypersetup{colorlinks=true,citecolor=blue,linkcolor=blue,urlcolor=blue}

\usepackage[mathlines]{lineno}

\usepackage{soul}

\usepackage{tikz,xcolor,hyperref}
\definecolor{lime}{HTML}{A6CE39}
\DeclareRobustCommand{\orcidicon}{%
	\begin{tikzpicture}
	\draw[lime, fill=lime] (0,0) 
	circle [radius=0.16] 
	node[white] {{\fontfamily{qag}\selectfont \tiny ID}};
	\draw[white, fill=white] (-0.0625,0.095) 
	circle [radius=0.007];
	\end{tikzpicture}
	\hspace{-2mm}
}

\foreach \x in {A, ..., Z}{%
	\expandafter\xdef\csname orcid\x\endcsname{\noexpand\href{https://orcid.org/\csname orcidauthor\x\endcsname}{\noexpand\orcidicon}}
}

\begin{document}
\title{Pseudoscalar Meson Parton Distributions Within Gauge-Invariant Nonlocal Chiral Quark Model}
\author{Parada~T.~P.~Hutauruk\orcidA{}}
\email{phutauruk@pknu.ac.kr}
\affiliation{Department of Physics, Pukyong National University (PKNU), Busan 48513, Republic of Korea}
\affiliation{Departemen Fisika, FMIPA, Universitas Indonesia, Depok 16424, Indonesia}

\date{\today}

\begin{abstract}
In this paper, I investigate the gluon distributions for the kaon and pion, as well as the improvement of the valence-quark distributions, in the framework of the gauge-invariant nonlocal chiral quark model (NL$\chi$QM), where the momentum dependence is taken into account. I then compute the gluon distributions for the kaon and pion that are dynamically generated from the splitting functions in the Dokshitzer–Gribov–Lipatov–Altarelli–Parisi (DGLAP) QCD evolution. In a comparison with the recent lattice QCD and JAM global analysis results, it is found that the results for the pion gluon distributions at $Q =$ 2 GeV, which is set based on the lattice QCD, have a good agreement with the recent lattice QCD data; this is followed up with the up valence-quark distribution of the pion results at $Q =$ 5.2 GeV in comparison with the reanalysis experimental data. 
The prediction for the kaon gluon distributions at $Q = 2$ GeV is consistent with the recent lattice QCD calculation.

\end{abstract}

\maketitle

\section{Introduction} \label{sec:intro}
The parton distribution function (PDF) is one of the important key tools for accessing the nonperturbative quantum chromodynamics (QCD)~\cite{Gross:2022hyw} domain of the hadron structure~\cite{Berger:1979du}, in addition to the electromagnetic form factor (EMFF), transverse momentum dependent (TMD), parton distribution amplitude (PDA), fragmentation function (FF), gravitational form factor (GFF), generalized transverse momentum dependent (GTMD), and generalized parton distribution (GPD). {Additionally, the} PDF will provide us with information {that is} crucial for further understanding {pseudoscalar} meson structure and the dynamical chiral symmetry breaking (DCSB) of nonperturbative QCD. {Note that, in the present work, we will limit our study only to the kaon and pion. Hereafter, the pseudoscalar meson refers to the kaon and pion}. In comparison to the nucleon PDFs, our knowledge and understanding of the pion and kaon PDFs is incomplete, in particular for the pion and kaon gluon distributions, because of the lack of a meson source target in experiments, leading to scarce meson data. These days, the circumstances are worse due to the current controversy on the power-law behavior of pion PDFs at large-$x$, which shows different predictions obtained among the theoretical models and analyses, when they are compared with the existing experimental data through the Drell--Yan process~{\cite{Conway:1989fs}.} 
 Therefore, 
 more data and theoretical studies are needed to resolve the current controversy and to arrive at a deeper understanding of the quark--gluon dynamics within light mesons.

Very recently, several suggestions and attempts to extract the pion and kaon PDFs and EMFFs from the experiments have been intensively discussed in the literature~\cite{Chavez:2021koz,Arrington:2021biu,Anderle:2021wcy}.
For example, in accessing the EMFFs of the pion data, the Sullivan process~\cite{Sullivan:1971kd} has provided a significantly larger value of transferred momentum $Q^2$ coverage~\cite{JeffersonLab:2008jve}. Analogous to the EMFFs of the pion, the previous authors also argued that extracting the kaon and pion PDFs is more feasible in the Sullivan process~\cite{AbdulKhalek:2021gbh}. Such a process is rather different from the ordinary reaction process used to extract the pion PDF, which was mostly obtained from the pion-induced Drell--Yan and $J/\psi$ production processes. Additionally, the COMPASS++/AMBER experiment at CERN~\cite{Adams:2018pwt} has been proposed to measure the cross-section of the pion--nucleus Drell--Yan process, where this will allow us to have more data in the large-$x$ regime and to access the pion gluon distribution functions (DFs). The availability of the kaon beam would allow us to have data for the kaon quark and gluon DFs. It is worth noting that the extraction of the kaon and pion gluon DFs is one of the focus programs of future experiments such as the electron--ion collider (EIC)~\cite{Arrington:2021biu}, electron--ion collider in China (EicC)~\cite{Anderle:2021wcy}, JPARC in Japan~\cite{Sawada:2016mao}, JLAB 22 GeV upgrade~\cite{Accardi:2023chb}, and COMPASS++/AMBER~\cite{Adams:2018pwt}. Therefore, the study of the present topic becomes relevant.

On the other hand, several theoretical studies have already been made to investigate the quark and gluon DFs for the pion and kaon~\cite{Hutauruk:2016sug,Hutauruk:2018zfk,Jia:2018ary,Kock:2020frx,Nam:2012vm,Hutauruk:2018qku,Hutauruk:2021kej,Hutauruk:2019ipp,Cui:2021mom,Albino:2022gzs,dePaula:2022pcb,Bourrely:2022mjf,Son:2024uet,Gifari:2024ssz,Han:2024yzj,Abidin:2019xwu}, as well as lattice QCD~\cite{Fan:2021bcr,Salas-Chavira:2021wui} and global QCD analysis~\cite{Novikov:2020snp,JeffersonLabAngularMomentumJAM:2022aix,Barry:2021osv}, to understand the dynamics of quarks and gluons~\cite{Aguilar:2020uqw}. Much impressive progress has been made so far in studying the kaon and pion gluon DFs. However, more theoretical studies and analyses with various approaches on gluon distribution certainly {are needed} to support the experimental physics program, since it is expected that the gluon contribution to the pion and kaon masses is considerable, which is approximately about (30--40)\% of the pion mass.

In this work, I investigate the quark and gluon DFs of the kaon and pion in the framework of the gauge-invariant nonlocal chiral quark model (NL$\chi$QM), considering momentum dependence. However, in this work, I {focus} on the gluon DFs of the kaon and pion. In computing the gluon DFs in the kaon and pion, I employ the next-to-leading-order Dokshitzer–Gribov–Lipatov–Altarelli–Parisi (NLO DGLAP) QCD evolution~\cite{Miyama:1995bd} to dynamically generate the gluon distributions at a higher scale $Q^2$. Next, I then compare the result with the existing {experimental} data~\cite{Conway:1989fs} and recent lattice QCD for the kaon and pion gluon DFs~\cite{Fan:2021bcr,Salas-Chavira:2021wui}. Overall, I find that the numerical results for the valence-quark and gluon DFs for the kaon and pion are consistent with the reanalysis data~\cite{Conway:1989fs} and recent lattice QCD~\cite{Fan:2021bcr,Salas-Chavira:2021wui}.

The present paper is organized as follows.
In Section~\ref{sec:nlchm}, I briefly begin by introducing the theoretical model formalism of the gauge-invariant NL$\chi$QM and show the expression for the valence-quark distributions for the kaon and pion after applying the gauge-invariant effective chiral action (EChA), which starts from a general expression of the PDFs. Section~\ref{sec:result} presents the numerical results for the kaon and pion gluon distributions. Finally, a summary and conclusion are provided in Section~\ref{sec:summary}. 

\section{Gauge-invariant nonlinear chiral quark model and PDF} \label{sec:nlchm}
Here, I present a general expression of the kaon and pion valence-quark DFs in terms of the two-point function (two-point correlation function) of local operators. I then decompose the general expression of the parton distribution into the kaon and pion PDFs in the gauge-invariant NL$\chi$QM. The generic expression for the twist-2 PDFs for the kaon and pion can be given by the following~\cite{Hutauruk:2023ccw}:
\begin{eqnarray}
    \label{eq1}
    f_{\phi} (x) &=& \frac{i}{4\pi} \int d\eta \exp[i(xp)\cdot(\eta n)] \Big< \phi (p) \Big| \bar{q}_f (\eta n) n\!\!\!/ q_f(0)\Big| \phi (p)\Big>.
\end{eqnarray}

{{The} momentum fraction of the struck quark in the light meson is defined by \mbox{$x = (k\cdot n)/(p\cdot n)$}} with $n$, $k$, and $p$ being the light-like vector, the parton momentum, and the light meson momentum, respectively. In the light-cone coordinate framework, one can be defined as $n\cdot v = v^+$.

Now I, in turn, present the framework of the gauge-invariant NL$\chi$QM. {In this NL$\chi$QM, the quark propagator is treated as momentum-dependent, which is so-called nonlocal~\cite{Diakonov:2002fq}. With this nonlocality, the NL$\chi$QM does not require a further regulator. This approach can provide quark confinement by generating the quark propagator without poles in real energies~\cite{Plant:1997jr}. This makes this model distinct from the Nambu--Jona--Lasinio (NJL) model, where the quark propagator is momentum-independent~\cite{Klevansky:1992qe}. It is worth noting that the NL$\chi$QM has a similar approach to the Instanton Liquid model~\cite{Shuryak:1988ff} and Dyson--Schwinger equation~\cite{Roberts:1994dr}}. The effective chiral action for the NL$\chi$QM can be \mbox{written as}
\begin{eqnarray}
    \label{eq:m1}
    \mathcal{S}_{\rm{eff}} [\phi,m_f,\mu] &=& -i \mathrm{Sp} \ln\Big[i\partial\!\!\!/ -\hat{m}_f - \sqrt{M_f(i\partial\!\!\!/)}U_5 \sqrt{M_f (i\partial\!\!\!/}\Big],
\end{eqnarray}
where $\mathrm{Sp}$ represents the functional trace of $\mathrm{Tr}_{f,c,\gamma} \langle \phi \mid \cdot \cdot \cdot \mid \phi \rangle$. The subscripts $c,f,\mathrm{and} \gamma$ represent the quark color, the quark flavor, and the Lorentz indices, respectively, and $\hat{m}_f$ is the current quark mass matrices $\mathrm{diag}(m_u,m_d,m_s)$. In this work, we consider the {SU(2)} isospin symmetry that is given by $m_u =m_d$, implying $M_u=M_d$, where the $M_f$ is the constituent quark mass for a given quark flavor $f$, which is momentum-dependent. The nonlinear term for the pseudoscalar meson fields, which is symbolized by $U_5$, can be expressed by
\begin{eqnarray}
    \label{eq:m2}
    U_5 &=& \exp \Big[ \frac{i\gamma_5 \lambda \cdot \phi}{\sqrt{2}F_{\phi}}\Big],~~~~~~~~~~~~~
    \lambda \cdot \phi  = \Bigg(~\begin{matrix} \frac{1}{2}\pi^0 + \frac{1}{\sqrt{6}}\eta & \pi^+ & K^+ \\ \pi^- & -\frac{1}{\sqrt{2}} \pi^0 + \frac{1}{\sqrt{6}} \eta & K^0 \\K^- & \bar{K}^0 & - \frac{2}{\sqrt{6}} \eta \end{matrix}~\Bigg),
\end{eqnarray}
where $F_\phi$ and $\lambda$ are the light meson weak-decay constants and the Gell--Mann matrix, respectively. The effective Lagrangian density for the quark--quark--meson ($q$-$q$-$\phi$) interaction vertex that is obtained from the effective chiral action is defined by
 $   \mathcal{L}_{qq\phi}^\mathrm{nonlocal} \simeq \frac{i}{F_\phi} \bar{q} \Big[ \sqrt{M_f(\partial)} \gamma_5 (\lambda \cdot \phi) \sqrt{M_f (\partial)} \Big]q$.
Taking off the momentum dependence of $M_f$ gives a positive constant value of constituent quark mass. 
Also, the local effective Lagrangian density for the pseudoscalar mesons can be straightforwardly obtained by 
    $\mathcal{L}_{qq\phi}^{\mathrm{local}} \simeq i g_{qq\phi} \bar{q} [\gamma_5 (\lambda \cdot \phi) ]q$ ,
where $g_{qq\phi}$ is the quark--quark meson coupling constant, which is a similar quantity obtained in the {NJL} model without momentum dependence. Preserving the gauge invariant of the EChA, we simply apply the minimal substitution ($\partial_\mu \rightarrow D_\mu = \partial_\mu - iV_\mu$ where $V$ is the local vector field) to the derivative of the quark in the kinetic part, and it gives 
\begin{eqnarray}
    \label{eq:m5}
    \mathcal{S}_{\mathrm{eff}}[\phi,m_f,V_\mu,\mu] &=& -\mathrm{Sp} \ln \Big[ iD\!\!\!\!/-\hat{m}_f - \sqrt{M_f (iD\!\!\!\!/)}U_5 \sqrt{M_f (iD\!\!\!\!/)}\Big].
\end{eqnarray}

Through the gauge-invariant EChA of Equation~(\ref{eq:m5}), I calculate the PDFs for the kaon and pion through a three-point functional derivative with respect to $\phi$ and $V$ with a $\delta$-function, that is, 
\begin{eqnarray}
    \label{eq:m6}
\frac{\delta^3\mathcal{S}_{\mathrm{eff}}[\phi,m_f,V_\mu,\mu]}{\delta \phi^\alpha (x) \delta \phi^\beta(y)\delta V_\mu (0)}\Bigg|_{\phi^{(\alpha, \beta)},V =0},
\end{eqnarray}
where the $\alpha$ and $\beta$ represent the isospin indices for the pseudoscalar mesons. I then perform the expansion on the nonlinear meson field $U_5$ in the effective chiral action up to the second order $\mathcal{O}(\phi^2)$. The PDF in the NL$\chi$QM is then expressed by

\begin{eqnarray}
    \label{eq:m7}
    f_{\phi} (x) &=& - \frac{iN_c}{2F_\phi^2} \int \frac{d^4k}{(2\phi)^4} \delta \left(k \cdot n - xp\cdot n\right) \nonumber \\
    &\times& \mathrm{Tr}_\gamma \Big[ \sqrt{M_b}\gamma_5 \sqrt{M_a} S_a n\!\!\!/ S_a \sqrt{M_a}\gamma_5\sqrt{M_b} S_b \nonumber \\
    &+& \left(\sqrt{M_b}\cdot n\right) \gamma_5 \sqrt{M_a}S_a\sqrt{M_a}\gamma_5\sqrt{M_b}S_b\nonumber \\
    &-& \sqrt{M_b}\gamma_5 \left(\sqrt{M_a} \cdot n\right)S_a \sqrt{M_a} \gamma_5 \sqrt{M_b}S_b\Big].
\end{eqnarray}
{Here}, the subscripts ($a,b$) stand for the quark flavors of the constituents in the meson with the corresponding momentum, $k_a = k$ and $k_b = k-p$, in the quark propagators. 
The second and third terms of Equation~(\ref{eq:m7}) contain, respectively, ($\sqrt{M_b}\cdot n$) and ($\sqrt{M_a}\cdot n$) and only appear when the momentum dependence of the effective quark mass is considered. 
These terms are so-called the \textit{{nonlocal}} or \textit{{derivative}} 
 interaction terms that are obtained from the functional derivative of the gauge-invariant effective chiral action with respect to $V_\mu$. The expression quark propagator for a given flavor $a$ is written as
\begin{eqnarray}
    \label{eq:m8}
    S_a (k_a) \equiv \frac{k_a\!\!\!\!\!\!/+ (m_a + M_a)}{k_a^2 -(m_a+M_a)^2 + i\epsilon} = \frac{k_a\!\!\!\!\!\!/+ \Tilde{M}_a}{k_a^2 - M_a^2 + i\epsilon}.
\end{eqnarray}
{Here,} $\Tilde{M}_a = (m_a + M_a)$ is the effective quark mass where $m_a$ is the current quark mass for flavor $a$ and the mass function $M_a$ is simply defined by
\begin{eqnarray}
    \label{eq:m9}
    M_a &=& M_0 \Big[\frac{\mu^2}{k_a^2-\mu^2 + i\epsilon}\Big]^2, \nonumber \\
    \sqrt{M_{a_\mu}} &=& -\sqrt{M_a} \frac{2k_{a_{\mu}}}{(k_a^2 -\mu^2 +i\epsilon)},
\end{eqnarray}
where $M_0$ is defined as the constituent quark mass at zero momentum. Thus, the expression of the PDF in Equation~(\ref{eq:m7}) can be rewritten in the light-cone coordinate using the light-cone variable, which is represented by
$    k\cdot n = k^+ = xP^+,~~~k^2 = k^+ k^- - k_\perp^2,~~~p^2 = m_\phi^2$ , and $ 
    k\cdot p = \frac{1}{2}(p^+ k^- + k^+ p^-)$.
With these light-cone variable definitions, we finally obtain the compact expression for the PDFs of the kaon and pion in the NL$\chi$QM, that is,\vspace{-6pt}
\begin{eqnarray}
    \label{eq:m11}
    f_\phi (x) &=& \frac{iN_c}{4F_\phi^2} \int \frac{dk^- d^2k_\perp}{(2\pi)^3} \Big[\mathcal{F}_L(k^-,k_\perp^2) + \mathcal{F}_{NL,a} (k^-,k_\perp^2) + \mathcal{F}_{NL,b} (k^-,k_\perp^2)\Big] + [x \leftrightarrow (1-x)], 
\end{eqnarray}
where $\mathcal{F}_L (k^-,k_\perp^2)$, $\mathcal{F}_{NL,a} (k^-,k_\perp^2)$, and $\mathcal{F}_{NL,b} (k^-,k_\perp^2)$ can be, respectively, defined by\vspace{-6pt}
\begin{eqnarray}
    \label{eq:m12}
    \mathcal{F}_{L} (k^-,k_{\perp}^2) &=& \frac{4p^+ \eta^2 D_b^4 \Big[\mathcal{N}_1 + \mathcal{N}_2 + \mathcal{N}_3 \Big]}{\mathcal{D}_1 \mathcal{D}_2}, ~~~
    \mathcal{F}_{NL,a} (k^-,k_{\perp}^2) = \frac{4p^+ \eta^2 \Big[\mathcal{N}_4 + \mathcal{N}_5 + \mathcal{N}_6 \Big]}{\mathcal{D}_1 \mathcal{D}_2}\Big[ \frac{x}{D_a^2}\Big],\nonumber \\
    \mathcal{F}_{NL,b} (k^-,k_{\perp}^2) &=& \frac{4p^+ \eta^2 \Big[\mathcal{N}_4 + \mathcal{N}_5 + \mathcal{N}_6 \Big]}{\mathcal{D}_1 \mathcal{D}_2} \Big[ \frac{x}{D_b^2}\Big],
\end{eqnarray}
where $\mathcal{N}_{1},\mathcal{N}_{2},\mathcal{N}_{3},\mathcal{N}_{4},\mathcal{N}_{5}$, and $ \mathcal{N}_{6}$ can be, respectively, defined by 

\begin{eqnarray}
\label{eq:m13}
    \mathcal{N}_1 &=& \left[ D_a^4 D_b^8((2-x)k_{\perp}^2 + (1-x)^2k^-p^+)\right], \nonumber \\
    \mathcal{N}_2 &=& \left[ 2(1-x)D_b^4(D_b^4m_b+\eta)(D_a^4m_a+ \eta)\right], \nonumber \\
    \mathcal{N}_3 &=& \left[xD_a^4 (D_b^4m_b + \eta)^2 \right], ~~~
    \mathcal{N}_4 = \left[ \eta (D_b^4m_b +\eta) \right],~~~
    \mathcal{N}_5 = \left[ D_a^4(m_a \eta)\right], \nonumber \\
    \mathcal{N}_6 &=& \Big[ D_a^4D_b^4(2k_\perp^2 + (1-2x)k^-p^+ + m_a m_b + x m_\phi^2)\Big],
\end{eqnarray}
and $\mathcal{D}_{1}$ and $\mathcal{D}_{2}$ are defined by
\begin{eqnarray}
\label{eq:m14}
    \mathcal{D}_1 &=& \left[ D_a^8 (\zeta_a -\alpha k^-) + 2m_a\eta D_a^4+ \eta^2 \right]_a,\nonumber \\
    \mathcal{D}_2 &=& \left[ D_b^8(\zeta_b-\beta k^- + \delta)+2m_b \eta D_b^4 + \eta^2 \right]_b^2,
\end{eqnarray}
and other variables are defined in terms of the light-cone variable by
\begin{eqnarray}
    \label{eq:m15}
    \alpha &=& xp^+,~~~~~~~~\beta = -(1-x)p^+,~~~
    \delta = -(1-x)m_\phi^2,~~~~~~\eta = M_0\mu^4, \nonumber \\
    \zeta_b &=& k_\perp^2 + m_b^2,~~~~~~D_a^2 =\gamma-\alpha k^-, ~~~
    D_b^2 = \gamma -\beta k^- +\delta,~~~~~~\gamma = k_\perp^2 + \mu^2, \nonumber \\
    \zeta_a &=& k_\perp^2 + m_a^2, ~~~
    p^2 = m_\phi^2 = p^+ p^- - p_\perp^2 \simeq p^+ p^-.
\end{eqnarray}
{Here}, it is worth noting that once the transverse momentum for the light meson is smaller compared to the longitudinal momentum, we can assume that $p_\perp \simeq 0$, implying $p^2 \simeq m_\phi^2$. Next, the PDF for the meson in Equation~(\ref{eq:m11}) can be solved by integrating out the $f_\phi (x)$ over $k^-$ and then numerically integrating over $k_\perp$ to obtain the final result for the PDF as a function of $x$. The PDFs for the light mesons should preserve the normalization condition, which gives
\begin{eqnarray}
    \label{eq:m16}
    \int_0^1 dx~f_{\phi} (x) = 1,
\end{eqnarray}
and the moments of the PDFs for the light meson can be calculated by
\begin{eqnarray}
\label{eq:m17}
    \langle x^m \rangle = \int_0^1 dx x^m f_\phi (x),
\end{eqnarray}
where $m = 0, 1, 2, \cdot \cdot \cdot$ is an integer.

Next, we present the QCD evolution of the PDFs for the light mesons using the DGLAP evolution. Thus, we can generate the gluon, quark, and sea-quark DFs. The quark (nonsinglet) DFs can be obtained by
\begin{eqnarray}
    \label{eq:m18}
    q_{\mathrm{NS}} (x) &=& q(x) - \bar{q} (x),
\end{eqnarray}
where the quark and antiquark DFs are, respectively, represented by $q(x)$ and $\bar{q}(x)$. In the DGLAP QCD evolution, the evolution of the nonsinglet quark DFs can be generated by {a convolution of the splitting function} with the quark distribution, which simply gives 
\begin{eqnarray}
\label{eq:m19}
    \frac{\partial q_{\mathrm{NS} (x,Q^2)}}{\partial \ln(Q^2)} &=& P_{qq} (x,\alpha_s(Q^2)) \bigotimes q_{\mathrm{NS} (x,Q^2)}. 
\end{eqnarray}
{The} product of the convolution between the splitting function and the nonsinglet quark distribution can be obtained as follows:

\begin{eqnarray}
    \label{eq:m20}
    P_{qq} \bigotimes Q_{\mathrm{NS}} = \int_x^1 \frac{dz}{x} P \Big( \frac{x}{z}\Big) q_{\mathrm{NS}} (z,Q^2).
\end{eqnarray}
{For} the singlet quark distribution, one has
\begin{eqnarray}
    \label{eq:m21}
    q_{\mathrm{S}} (x) = \sum q_{i}^+ = \sum_i q_i(x) + \bar{q}_i (x),
\end{eqnarray}
where the subscript of $i$ represents the quark flavor. The evolution of the singlet quark distribution can be given by
\begin{eqnarray}
    \label{eq:m22}
    \frac{\partial}{\partial \ln(Q^2)} \Bigg[\begin{matrix}
        q_{\mathrm{S}} (x,Q^2) \\ g(x,Q^2)
    \end{matrix} \Bigg] = \Bigg[ \begin{matrix}
        P_{qq} & P_{qg} \\
        P_{gq} & P_{gg}
    \end{matrix}\Bigg]\bigotimes \Bigg[ \begin{matrix}
        q_{\mathrm{S}}(x,Q^2) \\ g(x,Q^2)
    \end{matrix}\Bigg].
\end{eqnarray}
{In} Equation (\ref{eq:m22}), it is clearly shown that the gluon distribution is generated in the evolution of the singlet quark distribution of the DGLAP QCD evolution. The splitting functions can be perturbatively expanded in terms of $\alpha_s(Q^2)$, which gives $P(z, Q^2) = \left(\frac{\alpha_s}{2\pi} \right)P^{(0)} (z) + \left( \frac{\alpha_s}{2\pi}\right)^2 P^{(1)} (z) + \cdot \cdot \cdot $, where the first term of $P^{(0)}$ represents the leading order (LO), the second term of $P^{(1)}$ is the next-to-leading order (NLO), and the next terms of $(\cdot \cdot \cdot)$ are the next-to-next-to-leading order (NNLO), and so forth. Note that, in this work, we will focus on the NLO term, meaning that we evolve our PDF result up to the next-to-leading order. It is worth noting that the computation with the NLO and NNLO DGLAP QCD evolutions leads to almost unchanged valence-quark distributions, meaning that the NNLO has negligible effects on the DGLAP QCD evolution of the {valence}-quark DFs{, as concluded in Ref.~\cite{Lan:2019vui}}.

\section{Numerical result and Discussion}\label{sec:result}
In this section, the numerical results for the gluon DFs for the kaon and pion, as well as the valence-quark DFs, are presented. The model parameters of zero virtuality constituent quark mass $M_0 =$ 300 MeV and the model renormalization scale $\mu =$ 1 GeV used in the calculation are determined to preserve the PDF normalization condition and to reproduce the meson weak-decay constants. In this work, we use the kaon and pion weak-decay constants, respectively, $F_\pi =$ 93.2 MeV and $F_K =$ 113.4 MeV, as used in Ref.~\cite{Hutauruk:2023ccw}. The current quark masses for $m_u =m_d =$ 5 MeV and $m_s =$ 100 MeV are chosen in the present~work.

\begin{figure}[H]
    \centering
    \includegraphics[scale=1.1]{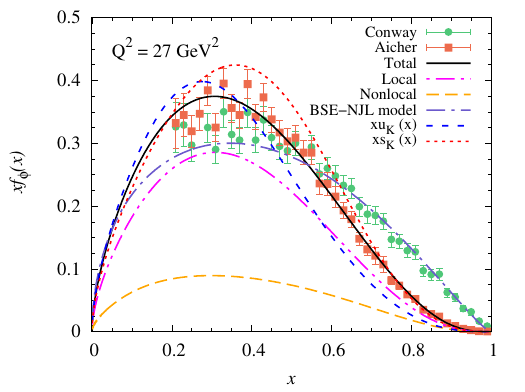}
    \caption{{Valence-quark} 
 DFs for the pion and kaon multiplying by the longitudinal momentum $x$ at $Q =$ 5.2 GeV that evolved from the initial model scale $Q_0 =$ 0.42 GeV. The solid, dashed, and dotted lines represent the up valence-quark DFs for the pion, the up valence-quark DFs for the kaon, and the strange valence-quark DFs for the kaon, respectively. The experimental data are taken from the muon-pair production experiment~\cite{Conway:1989fs} and the reanalysis experiment data~\cite{Aicher:2010cb}. The dash-dotted line represents the up valence-quark DFs for the pion that were calculated in the BSE-NJL model~\cite{Hutauruk:2021kej}, while the local and nonlocal contributions for the pion DFs are represented by the dot-dot-dashed and long dashed lines, respectively.}
    \label{fig1}
\end{figure}

Results for the valence-quark DFs for the pion and kaon at $Q =$ 5.2 GeV (a typical scale of the experimental data), to compare them with experimental data, are shown in Figure~\ref{fig1}. One shows that the pion's up valence-quark PDFs at $Q =$ 5.2 GeV fit remarkably well with the reanalysis data~\cite{Aicher:2010cb}. However, in comparison with the original empirical data in Ref.~\cite{Conway:1989fs}, the present result does not fit well with the data at around $x \gtrsim$ 0.6, but it is rather compatible with the data at around $x \simeq$ 0.45. It is worth noting that it has also been checked for different values of $Q_0$, as reported in Ref.~\cite{Nam:2012vm}; while not shown here, it was found that $Q_0 =$ 0.42 GeV fits the experimental data. 
I also show the result of the local (Long-dashed line) and nonlocal (Dot-dashed line) contributions for the pion's up valence-quark DFs at $Q =$ 5.2 GeV. The local and nonlocal contributions of the pion valence-quark DFs in the NL$\chi$QM model are shown, indicating that the preservation of gauge invariance is explicitly considered in the calculation to conserve the axial-vector current. 
Figure~\ref{fig1} also clearly shows that the contribution of the nonlocal DFs for the pion is significant enough to produce the experimental data. 

At $Q =$ 5.2 GeV, the up valence-quark carries fraction momentum relative to the pion momentum with 2$\langle x \rangle_\pi =$ 0.41. The nonlocal and local up valence-quarks at \mbox{$Q = $ 5.2 GeV} carrying fraction momentum relative to the pion momentum are about {0.11} and {0.31, respectively}. 
The total, local, and nonlocal contributions of the pion DFs at $Q =$ 5.2 GeV for higher moments are listed in Table~\ref{tab1}. 
In addition, I present the results of the large-$x$ behavior of the up valence-quark DFs for the pion, where the data are fit with a range of $x=[0.8,1.0]$ to the power-law form $(1-x)^{a}$. I found that the up valence-quark DFs for the pion at \mbox{$Q =$ 5.2 GeV} had a power-law behavior of $(1-x)^{1.85}$ at the endpoint of $x =$ 1. 
This is consistent with the result obtained in Ref.~\cite{Cui:2021mom}, where the momentum dependence is \mbox{also considered.}
\begin{table}[H]
	\begin{ruledtabular}
		\renewcommand{\arraystretch}{1.3}
		\caption{The moments of the pion valence-quark DFs at $Q =$ 5.2 GeV that evolved from \mbox{$Q_0 =$ 0.42 GeV.} The subscripts of $\mathrm{L}$ and $\mathrm{NL}$ stand for the linear and nonlinear contributions, respectively.}
		\label{tab1}
		\begin{tabular}{cccccccc}
		 \textbf{Types} & \boldmath{$Q$} & \boldmath{$n=1$} & \boldmath{$n=2$}  & \boldmath{$n=3$} & \boldmath{$n=4$} & \boldmath{$n=5$} & \boldmath{$n=6$}  \\  \hline
    Up & &  &  &  &  &  &  \\  \hline
		 $\langle x^n \rangle_{\pi}$ & 5.2 & 0.21 & 0.08 & 0.04 & 0.02 & 0.01 & 0.01 \\
   $\langle x^n \rangle^{\pi}_{\mathrm{L}}$ & 5.2 & 0.16 & 0.06 & 0.03 & 0.02 & 0.01 & 0.01 \\
   $\langle x^n \rangle^{\pi}_{\mathrm{NL}}$ & 5.2 & 0.05 & 0.02 & 0.01 & 0.01 & 0.004 & 0.002 \\ \hline
       Gluon & &  &  &  &  &  &  \\ \hline
        $\langle x^n \rangle^{\pi}_g$ & 5.2 & 0.62 & 0.06 & 0.01 & 0.004 & 0.002 & 0.001 \\
   $\langle x^n \rangle^{\pi}_{g,\mathrm{L}}$ & 5.2 & 0.52 & 0.01 & 0.004 & 0.001 & 0.0006 & 0.0003 \\
   $\langle x^n \rangle^{\pi}_{g,\mathrm{NL}}$ & 5.2 & 0.34 & 0.03 & 0.008 & 0.002 & 0.0009 & 0.0004 \\ 
		\end{tabular}
		\renewcommand{\arraystretch}{1}
	\end{ruledtabular}
\end{table}

Figure~\ref{fig1} also shows that the nonlocal contribution to the total pion DFs needs to be fitted to the data~\cite{Aicher:2010cb}. Using a similar procedure, the power-law behavior of the local and nonlocal DFs at $ Q=$5.2 GeV can be checked, and it was found that the local DFs have a power-law behavior of $(1-x)^{2.06}$ at large-$x$, while the nonlocal DFs for the pion at \mbox{$Q=$ 5.2 GeV} have a power-law behavior given by $(1-x)^{2.46}$.

Compared to other theory predictions, in contrast with the valence-quark DFs for the pion in the NC$\chi$QM model, the result of the pion valence-quark DFs in the BSE-NJL model~\cite{Hutauruk:2021kej} fits well with the original empirical data~\cite{Conway:1989fs}, which is consistent with the JAM QCD analysis result. The power-law behavior for the BSE-NJL model is given by $(1-x)^{1.23}$ at large-$x$, which is also consistent with the JAM QCD analysis result. It is worth noting that we have used the same power-law form and range of $x$ in determining the power-law behavior in the BSE-NJL model. More interestingly, here I recapitulate again, as shown in Figure~\ref{fig1}, that the pion quark DFs in the BSE-NJL model without momentum dependence are consistent with old experimental data~\cite{Conway:1989fs}, whereas the pion quark DFs in the NL$\chi$QM with momentum dependence are consistent with reanalysis data~\cite{Aicher:2010cb}. These results may provide a good explanation for the long-standing puzzle of large-$x$ power-law behavior.

In addition to power-law and moment results, I present the parameterized result for the pion valence-quark DFs---the local and nonlocal DFs contributions at $Q$ = 5.2 GeV, which are, respectively, given by
\begin{eqnarray}
  xu_\pi (x) &=& 0.93 x^{0.60} (1-x)^{2.39} (1+2.95 x^{0.95}),\\
xu^\pi_{\mathrm{L}} (x) &=& 0.70 x^{0.60} (1-x)^{2.58} (1+ 3.90 x^{1.04}), \\
xu^\pi_{\mathrm{NL}} (x) &=& 0.29 x^{0.63} (1-x)^{1.94} (1+ 0.95 x^{0.86}).
\end{eqnarray}

Now, I turn to presenting the results of the kaon up valence-quark DFs (dashed line) at $Q =$ 5.2 GeV that evolved from the initial scale $Q_0 =$ 0.42 GeV, as depicted in Figure~\ref{fig1}. The peak position of the kaon up valence-quark DFs is located at around $x \simeq$0.30, as clearly shown in Figure~\ref{fig1}. A similar indication is shown in the peak positions for the local and nonlocal contributions for the kaon. The power-law behavior for the up valence-quark DFs of the kaon at $Q =$ 5.2 GeV is given by $(1-x)^{2.30}$, while for the local and nonlocal DFs, the power-law behaviors are given by $(1-x)^{2.56}$ and $(1-x)^{2.87}$ at large-$x$. Here, I also provide the parameterized result for the kaon up valence-quark DFs as well as the local and nonlocal DFs at $Q$ = 5.2 GeV, which are given by
\begin{eqnarray}
xu_K (x) &=& 0.89 x^{0.58} (1-x)^{3.46} (1+ 7.90 x^{1.13}), \\
xu^K_{\mathrm{L}} (x) &=& 0.69 x^{0.59} (1-x)^{3.69} (1+ 9.56 x^{1.22}), \\
xu^K_{\mathrm{NL}} (x) &=& 0.20 x^{0.56} (1-x)^{2.87} (1+ 4.73 x^{0.88}).
\end{eqnarray}

Results for the kaon-antistrange valence-quark DFs at $Q =$ 5.2 GeV are also depicted in Figure~\ref{fig1}. Evolving the kaon-antistrange quark DFs to $Q =$ 5.2 GeV, it clearly shows that the kaon-up valence-quark DFs have smaller values in comparison with the kaon-antistrange quark DFs, as found in other theoretical studies. This is very well understood because the strange quark has a heavier mass than that of the light quark, indicating the strange quark carries more kaon--meson momentum than the light quark. The power-law behavior for the antistrange quark DFs is given by $(1-x)^{1.92}$, while the local and nonlocal DFs for the kaon have power-law behaviors, which are, respectively, given by $(1-x)^{2.13}$ and $(1-x)^{2.54}$. The parameterization forms for the antistrange DFs, local, and nonlocal DFs at $Q =$ 5.2 GeV are given by
\begin{eqnarray}
xs_K (x) &=& 1.10 x^{0.64} (1-x)^{3.06} (1+ 9.93 x^{1.60}), \\
xs^K_{\mathrm{L}} (x) &=& 0.82 x^{0.65} (1-x)^{3.19} (1+ 11.67 x^{1.67}), \\
xs^K_{\mathrm{NL}} (x) &=& 0.28 x^{0.63} (1-x)^{2.69} (1+ 6.05 x^{1.40}).
\end{eqnarray}

\begin{figure}[H]
    \centering
    \includegraphics[scale=1.1]{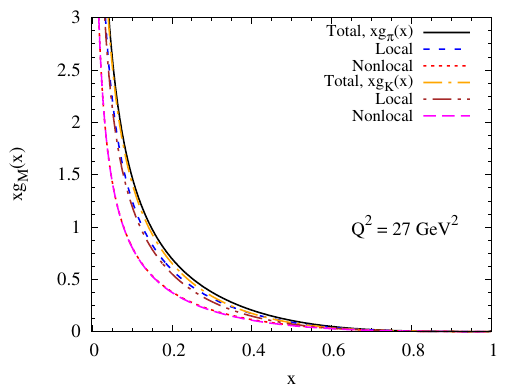}
    \caption{Gluon DFs for the pion and kaon at $Q =$ 5.2 GeV with contributions of the local and nonlocal DFs. The solid, dashed, and dotted lines represent the total gluon DFs for the pion, local, and nonlocal DFs, respectively. For the kaon, the total gluon DFs and the local and nonlocal DFs are given by dash-dotted, dot-dashed, and long-dashed lines, respectively.}
    \label{fig2}
\end{figure}

Numerical results for the gluon DFs for the pion and kaon include their local and nonlocal DFs at $Q =$ 5.2 GeV and are shown in Figure~\ref{fig2}. 
It clearly shows the significant contributions of the local and nonlocal gluon DFs to the total gluon DFs for the pion. Figure~\ref{fig2} also shows, similar to the gluon DFs for the pion, the nonlocal contribution of the gluon DFs to the total gluon DFs for the kaon, which is evolved from $Q_0 =$ 0.42 GeV, following the valence-quark DFs initial scale in Figure~\ref{fig1}. The local contribution of the gluon DFs for the kaon at $Q =$ 5.2 GeV is represented by a dash-dotted line. Unfortunately, there is no experimental data yet available for the pion and kaon gluon DFs. Hence, EIC, EicC, and COMPASS++/AMBER data are needed to confirm the result of this study. I also computed the total momentum carried by the gluon at $Q =$ 5.2 GeV. It was found that the gluon carries momentum relative to the total pion momentum that is about $\langle x \rangle_g^{\pi} =$ 0.62. The computation of the moments for the gluon DFs for the pion at $Q =$ 5.2 GeV is given in Table~\ref{tab1}, while for the kaon, the moment values are provided in Table~\ref{tab2}.

\vspace{-4pt}
\vspace{-6pt}
\begin{table}[H]
	\begin{ruledtabular}
		\renewcommand{\arraystretch}{1.3}
		\caption{The moments of the kaon valence-quark DFs at $Q =$ 5.2 GeV that evolved from \mbox{$Q_0 =$ 0.42 GeV.} Note that $\mathrm{L}$ and $\mathrm{NL}$ are the same as in Table~\ref{tab1}.}
		\label{tab2}
		\begin{tabular}{cccccccc}
		 \textbf{Types} & \boldmath{$Q$} & \boldmath{$n=1$} & \boldmath{$n=2$}  & \boldmath{$n=3$} & \boldmath{$n=4$} & \boldmath{$n=5$} & \boldmath{$n=6$}  \\ \hline
    Up & &  &  &  &  &  &  \\ \hline
		 $\langle x^n \rangle_{K}$ & 5.2 & 0.19 & 0.06 & 0.03 & 0.01 & 0.007 & 0.004 \\
   $\langle x^n \rangle^{K}_{\mathrm{L}}$ & 5.2 & 0.14 & 0.05 & 0.02 & 0.01 & 0.005 & 0.003 \\
   $\langle x^n \rangle^{\pi}_{\mathrm{NL}}$ & 5.2 & 0.05 & 0.02 & 0.01 & 0.004 & 0.002 & 0.001 \\ \hline
       Antistrange & &  &  &  &  &  &  \\ \hline
		 $\langle x^n \rangle_{K}$ & 5.2 & 0.23 & 0.09 & 0.04 & 0.02 & 0.01 & 0.01 \\
   $\langle x^n \rangle^{K}_{\mathrm{L}}$ & 5.2 & 0.17 & 0.06 & 0.03 & 0.02 & 0.01 & 0.01 \\
   $\langle x^n \rangle^{\pi}_{\mathrm{NL}}$ & 5.2 & 0.06 & 0.02 & 0.01 & 0.01 & 0.004 & 0.002 \\ \hline
       Gluon & &  &  &  &  &  &  \\ \hline
        $\langle x^n \rangle^{K}_g$ & 5.2 & 0.60 & 0.05 & 0.01 & 0.004 & 0.001 & 0.0005 \\
   $\langle x^n \rangle^{K}_{g,\mathrm{L}}$ & 5.2 & 0.50 & 0.04 & 0.01 & 0.003 & 0.001 & 0.0004 \\
   $\langle x^n \rangle^{\pi}_{g,\mathrm{NL}}$ & 5.2 & 0.34 & 0.03 & 0.01 & 0.002 & 0.0009 & 0.0004  \\
		\end{tabular}
		\renewcommand{\arraystretch}{1}
	\end{ruledtabular}
\end{table}

The power-law of the gluon DFs for the pion at $Q =$ 5.2 GeV behaves as $(1-x)^{3.06}$ at large-$x$. For the local and nonlocal DFs for the pion, their power-law behaviors are, respectively, given by $(1-x)^{3.11}$ and $(1-x)^{3.16}$. The gluon DF parametrizations at \mbox{$Q =$ 5.2 GeV} for the pion and kaon, as well as their local and nonlocal gluon DFs for the pion and kaon, are  given by
\begin{eqnarray}
    \label{eq:nr1}
    xg_\pi (x) &=& 1.47 x^{-0.42} (1-2.55\sqrt{x}+4.74x) (1-x)^{5.28},\\
xg^\pi_{\mathrm{L}} (x) &=& 1.24 x^{-0.42} (1-2.53\sqrt{x} + 4.72 x) (1-x)^{5.29}, \\
xg^\pi_{\mathrm{NL}} (x) &=& 0.78 x^{-0.42} (1-2.43\sqrt{x}+4.33 x) (1-x)^{5.01}, \\
 xg_K (x) &=& 1.45 x^{-0.42} (1-2.61\sqrt{x}+5.16x) (1-x)^{5.82},\\
xg^K_{\mathrm{L}} (x) &=& 1.21 x^{-0.42} (1-2.58\sqrt{x} + 5.08 x) (1-x)^{5.76}, \\
xg^K_{\mathrm{NL}} (x) &=& 0.79 x^{-0.42} (1-2.46\sqrt{x}+4.54 x) (1-x)^{5.26}.
\end{eqnarray}

Following the gluon DFs for the pion at $ Q=$5.2 GeV, here, we compute the large-$x$ power-law behavior for the kaon at $Q =$ 5.2 GeV. It is found that the gluon DFs for the kaon behave as $(1-x)^{3.17}$, while the local and nonlocal gluon DFs for the kaon behave as $(1-x)^{3.17}$ and $(1-x)^{3.21}$, respectively.

\begin{figure}[H]
\centering
    \includegraphics[scale=1.2]{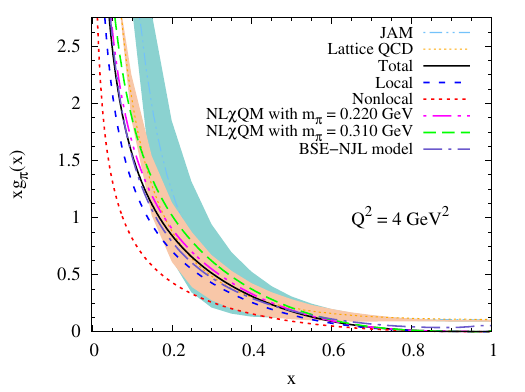}\\
    \includegraphics[scale=1.2]{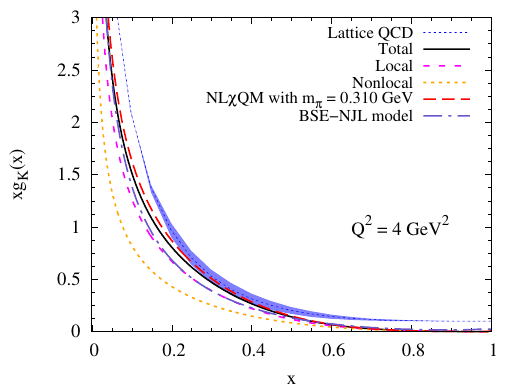}
    \caption{{Same} 
 as in Figure~\ref{fig2}, but gluon DFs for the kaon, in comparison with the lattice QCD data. Lattice QCD data is taken from Ref.~\cite{Salas-Chavira:2021wui}. }
    \label{fig5}
\end{figure}

To compare with the recent lattice QCD data~\cite{Fan:2021bcr}, we evolve the kaon and pion gluon DFs at $Q =$ 2 GeV. The result for the pion gluon DFs at $Q=$ 2 GeV, as adapted from Refs.~\cite{Fan:2021bcr,Aicher:2010cb,Sutton:1991ay}, shows good agreement with recent lattice data~\cite{Fan:2021bcr} and Jefferson Lab Angular Momentum (JAM) global QCD analysis~\cite{Barry:2021osv}, as clearly shown in the upper panel of Figure~\ref{fig5}. Note that the recent lattice QCD result for the pion gluon DFs is calculated at physical pion masses $m_\pi =$ 0.220 GeV and 0.310 GeV. Following the lattice QCD calculation, we then compute the gluon DFs of the pion at $Q =$ 2 GeV using similar values of the pion physical masses, and the results are shown with green long-dashed and magenta dash-dotted lines, which are comparable with the lattice QCD data.

In the upper panel of Figure~\ref{fig5}, we provide the nonlocal contribution of the gluon DFs of the pion at $Q =$ 2 GeV as well as the local contribution. The local contribution, which is approximately equivalent to the gluon DFs of the pion in the BSE-NJL model~\cite{Hutauruk:2021kej}, also shows good agreement with the lattice QCD data~\cite{Fan:2021bcr}. Note that the nonlocal contribution of the gluon DFs for the pion is relatively small, but it contributes significantly {to the total gluon distributions}. 

Results for the gluon DFs of the kaon at $Q =$ 2 GeV are shown in the lower panel of Figure~\ref{fig5}. It indicates that the gluon DFs of the kaon are consistent with the recent kaon lattice data~\cite{Salas-Chavira:2021wui} at $x \simeq$ 0.25 up to $x\simeq $ 0.50. Results for the local and nonlocal gluon DFs of the kaon at $Q=$ 2 GeV are shown in the lower panel of Figure~\ref{fig5}. The local gluon DFs of the kaon are quite similar to those obtained in the BSE-NJL model~\cite{Hutauruk:2021kej}. Nonlocal gluon DFs of the kaon have a smaller size but significantly contribute to the total gluon DFs of the kaon.

\begin{figure}[H]
\centering
    \includegraphics[scale=1.2]{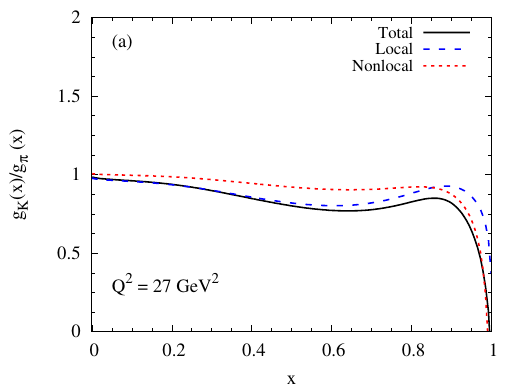}\\
    \includegraphics[scale=1.2]{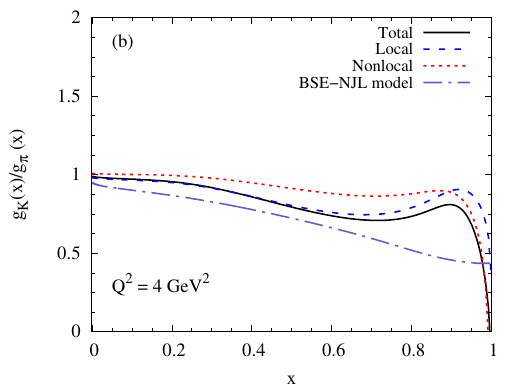}
    \caption{Ratios of the gluon DFs for the kaon and pion at (\textbf{a}) $Q =$ 5.2 GeV and (\textbf{b}) $Q = $ 2 GeV.}
    \label{fig6}
\end{figure}

To better understand the gluon dynamics inside the pion or the kaon, the comparison between the gluon DFs for the pion and the gluon DFs for the kaon is shown. Results are depicted in Figure~\ref{fig6} with $Q =$ 2 GeV and $Q =$ 5.2 GeV, where these values are adapted from the lattice QCD and empirical data, respectively.

Figure~\ref{fig6}a shows the ratios of the pion and kaon gluon DFs at $Q =$ 5.2 GeV that evolved using the same initial scales. It is found that the ratios decrease as the longitudinal momentum $x$ increases up to $x \simeq$ 0.7. However, different behaviors are shown between $ x \simeq$ 0.8 and $x =$ 1.0. The ratios slightly increase at $x \simeq$ 0.9 and decrease again at $x=$ 1.0 (endpoint). The results for the ratios of the pion and kaon gluon DFs are rather different from those obtained in the BSE-NJL model, which always decreases as the $x$ increases, as in Figure~\ref{fig6}b. This can be expected because of the {\textit{momentum-dependent}} contribution, which is absent in the BSE-NJL calculation. The decreasing behavior at around $x \simeq$1 is expected due to the transition region from soft to hard scales. Results for the ratio of the local and nonlocal gluon DFs to those of the pion are shown in Figure~\ref{fig6}a. Overall, one can conclude that the gluon DFs for the pion are larger than those for the kaon.

\vspace{-6pt}

A similar behavior is found in ratios of the pion and kaon gluon DFs at $Q =$ 2 GeV, as shown in Figure~\ref{fig6}b. This shows that the ratios of the gluon DFs of the kaon to that of the pion increase at $x \simeq$ 0.9 and decrease at $x \simeq$ 0.0 up to $x \simeq$ 0.7 and $x =$1.0. Decreasing the ratios indicates that the gluon DFs for the pion are larger than those in the kaon. This finding is consistent with other theoretical calculations of Ref.~\cite{Cui:2021mom} and references therein. However, the ratio for the pion and kaon in the BSE-NJL model at $Q =$ 2 GeV always decreases as the $x$ increases, meaning the gluon DFs of the pion are always larger than those for the kaon. This shows that the transition from the soft to hard scale is simpler than that found in the NL$\chi$QM, where momentum dependence is considered. Such behavior is also found in the Dyson--Schwinger equation model. However, to gain a deeper understanding of this behavior, it deserves further study.

\section{Summary} \label{sec:summary}
As a summary, in the present work, I have investigated the gluon DFs for the kaon and pion, as well as the valence-quark DFs in the framework of the gauge-invariant nonlocal chiral quark model (NL$\chi$QM), which considers momentum dependence. The gluon DFs were dynamically generated \textit{via} the splitting functions in the NLO DGLAP \mbox{QCD evolution. }

Results for the gluon DFs of the pion at $Q = $ 2 GeV, which was set to fit with the lattice QCD, show good agreement with the results of the recent lattice QCD~\cite{Fan:2021bcr} and JAM global QCD analysis~\cite{Barry:2021osv}. This prediction was followed by the up valence-quark DFs of the pion at $Q =$ 5.2 GeV that evolved from $Q_0 =$ 0.42 GeV in comparison with the reanalysis data~\cite{Aicher:2010cb}.

Interestingly, Figure~\ref{fig1} shows that the quark DFs of the pion that were computed in the BSE-NJL model without momentum dependence are consistent with old experimental data~\cite{Conway:1989fs}, whereas the pion quark DFs in the NL$\chi$QM with momentum dependence are consistent with reanalysis data~\cite{Aicher:2010cb}. These findings may provide a good explanation for the long-standing puzzle of large-$x$ power-law behavior.

The gluon DFs of the kaon at $Q =$ 2 GeV are found to be consistent with the recent kaon lattice QCD~\cite{Salas-Chavira:2021wui}. The prediction results for the total kaon gluon DFs at $Q =$ 5.2 GeV, which are taken based on experimental data, are also shown. Unfortunately, no data is available to perform comparisons with at the moment because of the scarcity of experimental data for the gluon. Ratios of the gluon DFs of the kaon and the pion at \mbox{$Q =$ 2} GeV show that the gluon DFs in the pion are larger than those in the kaon, which is consistent with the Dyson--Schwinger equation (DSE) results~\cite{Cui:2021mom}, considering momentum dependence. 

Finally, I have shown results for the up and antistrange valence-quark DFs of the kaon at $Q =$ 5.2 GeV, which are shown in Figure~\ref{fig1}. Unfortunately, like gluon DFs for the kaon and pion, no experimental data are now available for the valence-quark DFs for to compare the kaon results with. 
Furthermore, the parameterization of pion and kaon gluon and valence-quark DFs has been computed, which is useful for other calculations. 

 The findings of this work on the gluon DFs for the pion and kaon need to be confirmed by future modern facilities, such as via the electron-ion colliders (EIC)~\cite{Arrington:2021biu}, electron-ion colliders in China (EicC)~\cite{Anderle:2021wcy}, and COMPASS++/AMBER~\cite{Adams:2018pwt} experiments. 
 Also, the present results for the gluon and valence-quark DFs with local and nonlocal contributions would provide interesting guidance and information for the lattice QCD.

\section*{Funding}
This work was supported by the PUTI Q1 Research Grant from the University of Indonesia (UI) under contract No. NKB 442/UN2.RST/HKP.05.00/2024.


\section*{Acknowledgments}
P.T.P.H. thanks Huey-Wen Lin (Michigan State University) for providing them with their recent lattice QCD calculation results for the gluon DFs of the pion and kaon.

\section*{Conflicts of Interest}
The author declares no conflicts of interest.

\section*{Abbreviations}
The following abbreviations are used in this paper:

\begin{tabular}{@{}ll}
NL$\chi$QM & Nonlinear chiral quark model \\
QCD & Quantum chromodynamics\\
DGLAP & Dokshitzer–Gribov–Lipatov–Altarelli–Parisi\\
PDF & Parton distribution function \\
EMFF & Electromagnetic form factor \\
TMD & Transverse momentum dependent\\
PDA & Parton distribution amplitude \\
FF & Fragmentation function \\
GFF & Gravitational form factor \\
GTMD & Generalized transverse momentum dependent \\
GPD & Generalized parton distribution \\
DCSB & Dynamical chiral symmetry breaking \\
NLO& Next leading order\\
JAM & Jefferson Lab angular momentum \\
EIC & Electron-ion collider \\
EicC & Electron-ion collider in China \\
DF & Distribution function \\
ECha & Effective chiral action\\
NJL & Nambu--Jona--Lasinio \\
NNLO & Next-next leading order \\
BSE-NJL & Bethe--Salpeter equation--Nambu--Jona--Lasinio\\
\end{tabular}



\newpage 
\begin{thebibliography}{10}
\expandafter\ifx\csname url\endcsname\relax
  \def\url#1{\texttt{#1}}\fi
\expandafter\ifx\csname doi\endcsname\relax
  \def\doi#1{\burlalt{doi:#1}{http://dx.doi.org/#1}}\fi
\expandafter\ifx\csname urlprefix\endcsname\relax\def\urlprefix{URL }\fi
\expandafter\ifx\csname href\endcsname\relax
  \def\href#1#2{#2}\fi
\expandafter\ifx\csname burlalt\endcsname\relax
  \def\burlalt#1#2{\href{#2}{#1}}\fi
\bibitem{Gross:2022hyw}
Gross, F.; Klempt, E.; Brodsky, S.J.; Buras, A.J.; Burkert, V.D.; Heinrich, G.; Jakobs, K.; Meyer, C.A.; Orginos, K.; Strickland, M.; {et al.}
50 Years of Quantum Chromodynamics.
\emph{Eur. Phys. J. C} \textbf{2023}, \emph{83}, 1125.

\bibitem{Berger:1979du}
Berger, E.L.; Brodsky, S.J. Quark Structure Functions of Mesons and the Drell-Yan Process. \emph{Phys. Rev. Lett.} \textbf{1979}, \emph{42}, 940--944.

\bibitem{Conway:1989fs}
Conway, J.S.; Adolphsen, C.E.; Alexander, J.P.; Anderson, K.J.; Heinrich, J.G.; Pilcher, J.E.; Possoz, A.; Rosenberg, E.I.; Biino, C.; Greenhalgh, J.F.; {et al.} Experimental Study of Muon Pairs Produced by 252-GeV Pions on Tungsten.
\emph{Phys. Rev. D} \textbf{1989}, \emph{39}, 92--122.

\bibitem{Arrington:2021biu}
Arrington, J.; Gayoso, C.A.; Barry, P.C.; Berdnikov, V.; Binosi, D.; Chang, L.; Diefenthaler, M.; Ding, M.; Ent, R.; Frederico, T.; {et al.} Revealing the structure of light pseudoscalar mesons at the electron\textendash{}ion collider.
\emph{J. Phys. G} \textbf{2021}, \emph{48}, 075106.

\bibitem{Anderle:2021wcy}
Anderle, D.P.; Bertone, V.; Cao, X.; Chang, L.; Chang, N.; Chen, G.; Chen, X.; Chen, Z.; Cui, Z.; Dai, L.; {et al.} Electron-ion collider in China.
\emph{Front. Phys.} \textbf{2021}, \emph{16}, 64701.

\bibitem{Chavez:2021koz}
Ch\'avez, J.M.M.; Bertone, V.; Borrero, F.D.S.; Defurne, M.; Mezrag, C.; Moutarde, H.; Rodr\'\i{}guez-Quintero, J.; Segovia, J. Accessing the Pion 3D Structure at the US and China Electron-Ion Colliders.
\emph{Phys. Rev. Lett.} \textbf{2022}, \emph{128}, 202501.

\bibitem{Sullivan:1971kd}
{Sullivan, J.D. One pion exchange and deep inelastic} 
 electron-nucleon scattering. \emph{Phys. Rev. D} \textbf{1972}, \emph{5}, 1732--1737.


\bibitem{JeffersonLab:2008jve}
Huber, G.M.; Blok, H.P.; Horn, T.; Beise, E.J.; Gaskell, D.; Mack, D.J.; Tadevosyan, V.; Volmer, J.; Abbott, D.; Aniol, K.; {et al.} Charged pion form-factor between Q**2 = 0.60-GeV**2 and 2.45-GeV**2. II. Determination of, and results for, the pion form-factor.
\emph{Phys. Rev. C} \textbf{2008}, \emph{78}, 045203.

\bibitem{AbdulKhalek:2021gbh}
Khalek, R.A.; Accardi, A.; Adam, J.; Adamiak, D.; Akers, W.; Albaladejo, M.; Al-bataineh, A.; Alexeev, M.G.; Ameli, F.; Antonioli, P.; {et al.} Science Requirements and Detector Concepts for the Electron-Ion Collider: EIC Yellow Report.
\emph{Nucl. Phys. A} \textbf{2022}, \emph{1026}, 122447.

\bibitem{Adams:2018pwt}
Adams, B.; Aidala, C.A.; Akhunzyanov, R.; Alexeev, G.D.; Alexeev, M.G.; Amoroso, A.; Andrieux, V.; Anfimov, N.V.; Anosov, V.; Antoshkin, A.; {et al.} Letter of Intent: A New QCD facility at the M2 beam line of the CERN SPS (COMPASS++/AMBER). {\emph{arXiv} \textbf{2018},} 
 arXiv:1808.00848.

\bibitem{Sawada:2016mao}
Sawada, T.; Chang, W.C.; Kumano, S.; Peng, J.C.; Sawada, S.; Tanaka, K.
Accessing proton generalized parton distributions and pion distribution amplitudes with the exclusive pion-induced Drell-Yan process at J-PARC.
\emph{Phys. Rev. D} \textbf{2016}, \emph{93}, 114034.


\bibitem{Accardi:2023chb}
Accardi, A.; Achenbach, P.; Adhikari, D.; Afanasev, A.; Akondi, C.S.; Akopov, N.; Albaladejo, M.; Albataineh, H.; Albrecht, M.; Almeida-Zamora, B.; {et al.}
Strong interaction physics at the luminosity frontier with 22 GeV electrons at Jefferson Lab.
\emph{Eur. Phys. J. A} \textbf{2024}, \emph{60}, 173.

\bibitem{Hutauruk:2016sug}
Hutauruk, P.T.P.; Cloet, I.C.; Thomas, A.W. Flavor dependence of the pion and kaon form factors and parton distribution functions.
\emph{Phys. Rev. C} \textbf{2016}, \emph{94}, 035201.

\bibitem{Hutauruk:2018zfk}
Hutauruk, P.T.P.; Bentz, W.; Clo\"et, I.C.; Thomas, A.W.
Charge Symmetry Breaking Effects in Pion and Kaon Structure.
\emph{Phys. Rev. C} \textbf{2018}, \emph{97}, 055210.

\bibitem{Jia:2018ary}
Jia, S.; Vary, J.P. Basis light front quantization for the charged light mesons with color singlet Nambu\textendash{}Jona-Lasinio interactions.
\emph{Phys. Rev. C} \textbf{2019}, \emph{99}, 035206.


\bibitem{Kock:2020frx}
Kock, A.; Liu, Y.; Zahed, I. Pion and kaon parton distributions in the QCD instanton vacuum.
\emph{Phys. Rev. D} \textbf{2020}, \emph{102}, 014039.


\bibitem{Nam:2012vm}
Nam, S.i. Parton-distribution functions for the pion and kaon in the gauge-invariant nonlocal chiral-quark model.
\emph{Phys. Rev. D} \textbf{2012}, \emph{86}, 074005.


\bibitem{Hutauruk:2018qku}
Hutauruk, P.T.P.; Oh, Y.; Tsushima, K.
Electroweak properties of pions in a nuclear medium.
\emph{Phys. Rev. C} \textbf{2019}, \emph{99}, 015202.


\bibitem{Hutauruk:2021kej}
Hutauruk, P.T.P.; Nam, S.i. Gluon and valence quark distributions for the pion and kaon in nuclear matter.
\emph{Phys. Rev. D} \textbf{2022}, \emph{105},~3.


\bibitem{Hutauruk:2019ipp}
Hutauruk, P.T.P.; Cobos-Mart\'\i{}nez, J.J.; Oh, Y.; Tsushima, K.
Valence-quark distributions of pions and kaons in a nuclear medium. \emph{Phys. Rev. D} \textbf{2019}, \emph{100}, 094011.


\bibitem{Cui:2021mom}
Cui, Z.F.; Ding, M.; Morgado, J.M.; Raya, K.; Binosi, D.; Chang, L.; Papavassiliou, J.; Roberts, C.D.; Rodr\'\i{}guez-Quintero, J.; Schmidt, S.M. Concerning pion parton distributions.
\emph{Eur. Phys. J. A} \textbf{2022}, \emph{58}, 10.


\bibitem{Albino:2022gzs}
Albino, L.; Higuera-Angulo, I.M.; Raya, K.; Bashir, A. Pseudoscalar mesons: Light front wave functions, GPDs, and PDFs.
\emph{Phys. Rev. D} \textbf{2022}, \emph{106}, 034003.


\bibitem{dePaula:2022pcb}
de Paula, W.; Ydrefors, E.; Alvarenga, J.H.N.; Frederico, T.; Salm\`e, G. Parton distribution function in a pion with Minkowskian dynamics.
\emph{Phys. Rev. D} \textbf{2022}, \emph{105}, L071505.

\bibitem{Bourrely:2022mjf}
Bourrely, C.; Chang, W.C.; Peng, J.C. Pion Partonic Distributions in the Statistical Model from Pion-induced Drell-Yan and $J/\Psi$ Production Data.
\emph{Phys. Rev. D} \textbf{2022}, \emph{105}, 076018.


\bibitem{Son:2024uet}
Son, H.D.; Hutauruk, P.T.P.
Generalized parton distributions of the kaon and pion within the nonlocal chiral quark model.
\emph{Phys. Rev. D} \textbf{2025}, \emph{111}, 5.


\bibitem{Gifari:2024ssz}
Gifari, G.; Hutauruk, P.T.P.; Mart, T.
Nuclear medium meson structures from the Schwinger proper-time Nambu\textendash{}Jona-Lasinio model.
\emph{Phys. Rev. D} \textbf{2024}, \emph{110}, 1.


\bibitem{Han:2024yzj}
Han, C.; Kou, W.; Wang, R.; Chen, X.
Gluon distribution and mass decomposition of the pion and kaon.
\emph{Eur. Phys. J. C} \textbf{2024}, \emph{84},~389.


\bibitem{Abidin:2019xwu}
Abidin, Z.; Hutauruk, P.T.P.
Kaon form factor in holographic QCD.
\emph{Phys. Rev. D} \textbf{2019}, \emph{100}, 054026.


\bibitem{Fan:2021bcr}
Fan, Z.; Lin, H.W. Gluon parton distribution of the pion from lattice QCD.
\emph{Phys. Lett. B} \textbf{2021}, \emph{823}, {136778.} 


\bibitem{Salas-Chavira:2021wui}
Salas-Chavira, A.; Fan, Z.; Lin, H.W. First glimpse into the kaon gluon parton distribution using lattice QCD.
\emph{Phys. Rev. D} \textbf{2022}, \emph{106}, 094510.


\bibitem{Novikov:2020snp}
Novikov, I.; Abdolmaleki, H.; Britzger, D.; Cooper-Sarkar, A.; Giuli, F.; Glazov, A.; Kusina, A.; Luszczak, A.; Olness, F.; Starovoitov, P.; {et al.} Parton Distribution Functions of the Charged Pion Within The xFitter Framework.
\emph{Phys. Rev. D} \textbf{2020}, \emph{102}, 014040.


\bibitem{JeffersonLabAngularMomentumJAM:2022aix}
Barry, P.C.; Egerer, C.; Karpie, J.; Melnitchouk, W.; Monahan, C.; Orginos, K.; Qiu, J.W.; Richards, D.; Sato, N.; Sufian, R.S.; {et al.} 
Complementarity of experimental and lattice QCD data on pion parton distributions.
\emph{Phys. Rev. D} \textbf{2022}, \emph{105}, 114051.


\bibitem{Barry:2021osv}
Barry, P.C.; Ji, C.R.; Sato, N.; Melnitchouk, W. Global QCD Analysis of Pion Parton Distributions with Threshold Resummation.
\emph{Phys. Rev. Lett.} \textbf{2021}, \emph{127}, 232001.


\bibitem{Aguilar:2020uqw}
Aguilar, A.C.; Ferreira, M.N.; Papavassiliou, J. Gluon dynamics from an ordinary differential equation.
\emph{Eur. Phys. J. C} \textbf{2021}, \emph{81}, 54.


\bibitem{Miyama:1995bd}
Miyama, M.; Kumano, S. Numerical solution of $Q^2$ evolution equations in a brute force method.
\emph{Comput. Phys. Commun.} \textbf{1996}, \emph{94}, 185--215.


\bibitem{Hutauruk:2023ccw}
Hutauruk, P.T.P.; Nam, S.i.
Updated analyses of gluon distribution functions for the pion and kaon from the gauge-invariant nonlocal chiral quark model.
\emph{Phys. Rev. D} \textbf{2024}, \emph{109}, 054040.


\bibitem{Diakonov:2002fq}
Diakonov, D. Instantons at work.
\emph{Prog. Part. Nucl. Phys.} \textbf{2003}, \emph{51}, 173--222.


\bibitem{Plant:1997jr}
Plant, R.S.; Birse, M.C. Meson properties in an extended nonlocal NJL model.
\emph{Nucl. Phys. A} \textbf{1998}, \emph{628}, 607--644.


\bibitem{Klevansky:1992qe}
Klevansky, S.P. The Nambu-Jona-Lasinio model of quantum chromodynamics.
\emph{Rev. Mod. Phys.} \textbf{1992}, \emph{64}, 649--708.


\bibitem{Shuryak:1988ff}
Shuryak, E.V. The instanton liquid.
\emph{Z. Phys. C} \textbf{1988}, \emph{38}, 165--172.


\bibitem{Roberts:1994dr}
Roberts, C.D.; Williams, A.G. Dyson-Schwinger equations and their application to hadronic physics.
\emph{Prog. Part. Nucl. Phys.} \textbf{1994}, \emph{33}, 477--575.


\bibitem{Lan:2019vui}
Lan, J.; Mondal, C.; Jia, S.; Zhao, X.; Vary, J.P. Parton Distribution Functions from a Light Front Hamiltonian and QCD Evolution for Light Mesons.
\emph{Phys. Rev. Lett.} \textbf{2019}, \emph{122}, 172001.


\bibitem{Aicher:2010cb}
Aicher, M.; Schafer, A.; Vogelsang, W. Soft-gluon resummation and the valence parton distribution function of the pion.
\emph{Phys. Rev. Lett.} \textbf{2010}, \emph{105}, 252003.


\bibitem{Sutton:1991ay}
Sutton, P.J.; Martin, A.D.; Roberts, R.G.; Stirling, W.J. Parton distributions for the pion extracted from Drell-Yan and prompt photon experiments.
\emph{Phys. Rev. D} \textbf{1992}, \emph{45}, 2349--2359.

\end{thebibliography}
\end{document}